\documentclass[11pt,a4,twoside]{article}
\usepackage[dvips,a4paper,left=2.5cm,right=2.5cm,top=2.5cm,bottom=2.5cm,headsep=1em]{geometry}
\usepackage{fancyhdr}
\usepackage{titlesec}
\usepackage[latin1]{inputenc}
\usepackage{amsmath,amsfonts}
\usepackage{rotating}
\usepackage[font={small}]{caption}
\usepackage{natbib}
\usepackage{lmodern,slantsc}
\usepackage{multirow}
\usepackage{chngcntr}
\usepackage{placeins}
\usepackage{caption}
\usepackage[breaklinks,
colorlinks=true, linkcolor=blue, citecolor=black, urlcolor=blue,
pdfborder={0 0 0}, pdfpagelabels]{hyperref}
\usepackage{tikz}
\def\vx{\mathbf x}

\def\vr{\mathbf r}

\def\v0{\boldsymbol{0}}

\def\veta{\boldsymbol{\eta}}

\newlength{\FigureHeight}
\newlength{\FigureHeightHalf}
\newcommand{\FigureXYLabel}[5]
{\settoheight{\FigureHeight}{#1}
\setlength{\FigureHeightHalf}{0.5\FigureHeight}
\begin{center}
\raisebox{\FigureHeightHalf}{\makebox{#4\makebox[#5]{}}}
#1\\
\vspace{#3}
#2\\
\end{center}}
\numberwithin{equation}{section}
\titleformat{\section}
{\large\bfseries}{\thetitle.}{0.5em}{}
\titleformat{\subsection}
{\normalfont\itshape\filcenter}{\normalfont\thetitle.}{0.5em}{}
\begin{document}

\title{On the use of applying Lie-group symmetry analysis to\\ turbulent channel
flow with streamwise rotation\\ {\Large\emph{A comment on the
article by Oberlack et al. (2006)}}}
\author{Michael Frewer$\,^1$\thanks{Email address for correspondence:
frewer.science@gmail.com}$\,\,$ \&$\,$ George Khujadze$\,^2$\\ \\
\small $^1$ Tr\"ubnerstra{\ss}e 42, 69121 Heidelberg, Germany\\
\small $^2$ Chair of Fluid Mechanics, Universit\"at Siegen, 57068
Siegen, Germany}

\date{{\small\today}}
\clearpage \maketitle \thispagestyle{empty}

\vspace{-1.5em}\begin{abstract}

\noindent The study by
\href{https://www.cambridge.org/core/journals/journal-of-fluid-mechanics/article/group-analysis-direct-numerical-simulation-and-modelling-of-a-turbulent-channel-flow-with-streamwise-rotation/3C2E60B364749F771719A592FEE17637}{Oberlack
{\it et al.}~(2006)} consists of two main parts: a direct
numerical simulation (DNS) of a turbulent plane channel flow with
streamwise rotation and a preceding Lie-group symmetry analysis on
the two-point correlation equation (TPC) to analytically predict
the scaling of the mean velocity profiles for different rotation
rates. We will only comment on the latter part, since the DNS
result obtained in the former part has already been commented on
by \cite{Recktenwald09}, stating that the observed mismatch
between DNS and their performed experiment is possibly due to the
prescription of periodic boundary conditions on a too small
computational domain in the spanwise direction.

By revisiting the group analysis part in
\href{https://www.cambridge.org/core/journals/journal-of-fluid-mechanics/article/group-analysis-direct-numerical-simulation-and-modelling-of-a-turbulent-channel-flow-with-streamwise-rotation/3C2E60B364749F771719A592FEE17637}{Oberlack
{\it et al.}~(2006)}, we will generate more natural scaling laws
describing better the mean velocity profiles than the ones
proposed. However, due to the statistical closure problem of
turbulence, this improvement is illusive. As we will demonstrate,
any arbitrary invariant scaling law for the mean velocity profiles
can be generated consistent to any higher order in the velocity
correlations. This problem of arbitrariness in invariant scaling
persists even if we would formally consider the infinite
statistical hierarchy of all multi-point correlation equations.
The closure problem of turbulence simply cannot be circumvented by
just employing the method of Lie-group symmetry analysis alone: as
the statistical equations are unclosed, so are their
symmetries$\hspace{.15mm}$! Hence, an\linebreak {\it a~priori}
prediction as how turbulence scales is thus not possible. Only
{\it a posteriori} by anticipating what to expect from numerical
or experimental data the adequate invariant scaling law\linebreak
can be generated through an iterative trial-and-error process.
Finally, apart from this issue, also several inconsistencies and
incorrect statements to be found in
\href{https://www.cambridge.org/core/journals/journal-of-fluid-mechanics/article/group-analysis-direct-numerical-simulation-and-modelling-of-a-turbulent-channel-flow-with-streamwise-rotation/3C2E60B364749F771719A592FEE17637}{Oberlack
{\it et al.}~(2006)} will~be~pointed out.

\vspace{0.5em}\noindent{\footnotesize{\bf Keywords:} {\it
Symmetries, Lie Groups, Invariant Scaling Laws, Turbulence,
Channel Flow, Rotation, Statistical
Mechanics, Two-point Correlation Equation, Closure Problem of Turbulence}}$\,$;\\
{\footnotesize{\bf PACS:} 47.10.-g, 47.27.-i, 47.85.-g, 05.20.-y,
02.20.-a, 02.50.-r}
\end{abstract}

\pagenumbering{arabic}\setcounter{page}{1}

\section{Lie-group symmetry analysis and the closure problem of turbulence\label{S1}}

The object under investigation in \cite{Oberlack06} for group
analysis is the {\it inviscid} $(\nu=0)$ two-point correlation
equation (TPC) [Eq.$\,$(2.14)] to generate invariant scaling laws
for large-scale quantities, such as mean velocities and Reynolds
stresses. Small scale quantities, such as the dissipation, are not
captured by this investigation. As said, ``the basis for this
assumption is the fact that, to leading order only, viscosity has
no effect as $Re\to\infty$. Viscosity only affects the small
scales of $\mathcal{O}(\eta)$ where $\eta$ is the Kolmogorov
length scale. Hence neglecting viscosity is only valid for $|\vr|
> \eta$ [the
large scales]" \cite[p.$\,$388]{Oberlack06}. Following this
strategy, as also cited from \cite{Oberlack02B}, \cite{Oberlack03}
or \cite{Khujadze04}, one may also take the one-point limit within
this TPC [Eq.$\,$(2.14)] to obtain the inviscid $(\nu=0)$ Reynolds
transport equations [Eqs.(2.1a)-(2.1c)], being valid then only
``in regions sufficiently far from solid walls, [where] the
viscous terms may be neglected to leading order, and the balance
is dominated by the pressure and the turbulent stresses"
\cite[p.$\,$385]{Oberlack06}.\footnote[2]{Note that by taking the
one-point limit $\vr\to\boldsymbol{0}$ within the inviscid TPC
[Eq.$\,$(2.14)], being itself only valid for the large scales
$|\vr|>\eta$ where $\eta\neq 0$ is the Kolmogorov length scale, it
is claimed that only an error of the order
$\mathcal{O}(Re^{-1/2})$ is made which becomes negligibly small if
the Reynolds number is large enough; see e.g. the statement made
in \cite{Khujadze04} [p.$\,$399].\label{footnote-p2}}

The symmetry analysis on Eq.$\,$(2.14) was performed as a group
classification in the way at looking which profiles for the mean
velocities are admitted under a certain given set of symmetries
for the higher-order moments including their coordinates. For the
details to be discussed below, we will present the key results
here again: If the mean velocity profiles $\bar{u}_1$ and
$\bar{u}_3$ satisfy the constraint equations
[Eqs.$\,$(2.18a)-(2.18b)]
\begin{equation}
\left.
\begin{aligned}
\left[a_1x_2+a_5\right]\frac{d\bar{u}_1(x_2)}{dx_2}-a_1\bar{u}_1(x_2)=c_1,\\
\left[a_1x_2+a_5\right]\frac{d\bar{u}_3(x_2)}{dx_2}-a_1\bar{u}_3(x_2)=c_3,
\end{aligned}
~~~\right\}\label{160915:0842}
\end{equation}
then the inviscid TPC system of equations [Eq.$\,$(2.14)]
\begin{equation}
\left.
\begin{aligned}
& \frac{\partial R_{2j}}{\partial x_2}-\frac{\partial
R_{ij}}{\partial r_i}=0,\quad\;\; \frac{\partial R_{ij}}{\partial
r_j}=0,\quad\;\;\frac{\partial\overline{pu_j}}{\partial
r_j}=0,\quad\;\;\frac{\partial\overline{u_2p}}{\partial x_2}-
\frac{\partial\overline{u_ip}}{\partial r_i}=0,\\[0.0em]
&\hspace{2.75cm} \frac{\partial T_{2j}^{(1)}}{\partial
x_2}-\frac{\partial T_{ij}^{(1)}}{\partial r_i}=0,\quad\;\;
\frac{\partial
T_{ij}^{(2)}}{\partial r_j}=0,\\[0.5em]
0=&
-R_{2j}\delta_{i1}\frac{d\bar{u}_1(x_2)}{dx_2}-R_{2j}\delta_{i3}\frac{d\bar{u}_3(x_2)}{dx_2}
-R_{i2}\delta_{j1}\frac{d\bar{u}_1(x_2+r_2)}{d(x_2+r_2)}
-R_{i2}\delta_{j3}\frac{d\bar{u}_3(x_2+r_2)}{d(x_2+r_2)}\\
& -\left[\bar{u}_1(x_2+r_2)-\bar{u}_1(x_2)\right]\frac{\partial
R_{ij}}{\partial r_1}
-\left[\bar{u}_3(x_2+r_2)-\bar{u}_3(x_2)\right]\frac{\partial
R_{ij}}{\partial r_3}\\
&
-\frac{1}{\rho}\left[\delta_{i2}\frac{\partial\overline{pu_j}}{\partial
x_2}-\frac{\partial\overline{pu_j}}{\partial
r_i}+\frac{\partial\overline{u_ip}}{\partial r_j}\right] -
T^{(1)}_{ij}-T^{(2)}_{ij}
-2\big[\epsilon_{1li}R_{lj}+\epsilon_{1lj}R_{il}\big],
\end{aligned}
~~~\right\}\label{160915:0858}
\end{equation}
admits, when written in its infinitesimal generator form, the
following four-parametric Lie-point symmetry group
[Eqs.$\,$(2.16)-(2.17)]\footnote[3]{The notation
$\mathsf{S}_1^{(1.1)}$ clarifies that the symmetry $\mathsf{S}_1$
\eqref{160915:0840} is connected to the constraint
\eqref{160915:0842}. In this respect it is also important to note
that since the system \eqref{160915:0858} is unclosed, all
admitted invariant transformations can only be regarded in the
weak sense as equivalence transformations, and not as true
symmetry transformations in the strong sense. For more details, we
refer e.g. to \cite{Frewer14.1} and the references therein. In the
following, however, we will continue to call them imprecisely as
``symmetries", like it was also done in \cite{Oberlack06}.}
\begin{align}
\!\!\!\mathsf{S}_1^{(1.1)}:&\;\;\; \xi_{r_1}=a_1r_1+a_2,\;\;\;
\xi_{r_2}=a_1r_2,\;\;\;
\xi_{r_3}=a_1r_3+a_4,\;\;\;\xi_{x_2}=a_1x_2+a_5,\nonumber\\
&\;\;\,
\eta_{R_{ij}}=2a_1R_{ij},\;\;\,\eta_{\overline{pu_i}}=3a_1\overline{pu_i},\;\;\,
\eta_{\overline{u_ip}}=3a_1\overline{u_ip},\;\;\,
\eta_{T^{{\scriptscriptstyle
(1)}}_{ij}}=2a_1T_{ij}^{{\scriptscriptstyle (1)}}, \;\;\,
\eta_{T^{(2)}_{ij}}=2a_1T_{ij}^{{\scriptscriptstyle (2)}},
\label{160915:0840}
\end{align}
and vice versa. Note that in \eqref{160915:0858} we included all
continuity conditions [Eq.$\,$2.5] consistently up to the highest
(unclosed) order, where we denoted these moments collectively
by~$T_{ij}$, in particular~as
\begin{equation}
T^{(1)}_{ij}=\frac{\partial R_{i(jk)}}{\partial r_k},\quad\;\;
T^{(2)}_{ij}=\frac{\partial R_{(i2)j}}{\partial
x_2}-\frac{\partial R_{(ik)j}}{\partial r_k}.\label{160920:0859}
\end{equation}
To simplify notation, we also suppressed the overall tilde-symbol
used in \cite{Oberlack06} to denote the re-scaling transformation
[Eqs.$\,$(2.13a)-(2.13c)] relative to the rotation
parameter~$\Omega$. To ensure that in the following no ambiguity
arises between the different notations, we will continually point
out every time when this transformation back  to the original
variables is needed  or performed.\pagebreak[4]

The above result put forward in \cite{Oberlack06}, however, gives
the misleading impression now that \eqref{160915:0840} is the only
symmetry that can be connected to the mean velocity
constraint~\eqref{160915:0842}. But by far this is not the case.
Instead, when performing a systematic symmetry analysis assisted,
e.g., by the Maple package DESOLV-II \citep{Vu12}, one obtains an
infinite-dimensional Lie-algebra involving arbitrary functions for
the dependent variables. For example, the following symmetry
\begin{align}
\mathsf{S}_2^{(1.1)}:&\;\;\; \xi_{r_1}=\alpha_1r_1+a_2,\;\;\;
\xi_{r_2}=\alpha_1r_2,\;\;\; \xi_{r_3}=\alpha_1r_3+a_4,
\;\;\;\xi_{x_2}=\alpha_1x_2+a_5,\nonumber\\[0.1em]
&\;\;\; \eta_{R_{11}}=\alpha_2R_{11}+f_{11}(x_2,r_2,r_3),\;\;\;
\eta_{R_{12}}=\alpha_2R_{12}+f_{12}(x_2,r_3),\nonumber\\[0.1em]
&\;\;\; \eta_{R_{13}}=\alpha_2R_{13}+f_{13}(x_2,r_2),\;\;\;
\eta_{R_{21}}=\alpha_2R_{21}+f_{21}(r_3),\;\;\;
\eta_{R_{22}}=\alpha_2R_{22}+f_{22}(r_1,r_3),\nonumber\\[0.1em]
&\;\;\;\eta_{R_{23}}=\alpha_2R_{23}+f_{23}(r_1),
\;\;\;\eta_{R_{31}}=\alpha_2R_{31}+f_{31}(x_2,r_2),\;\;\;
\eta_{R_{32}}=\alpha_2R_{32}+f_{32}(x_2,r_1),\nonumber\\[0.1em]
&\;\;\; \eta_{R_{33}}=\alpha_2R_{33}+f_{33}(x_2,r_1,r_2),\;\;\;
\eta_{\overline{pu_1}}=(\alpha_1+\alpha_2)\overline{pu_1}+g_1(x_2,r_2,r_3),\nonumber\\[0.1em]
&\;\;\;
\eta_{\overline{pu_2}}=(\alpha_1+\alpha_2)\overline{pu_2}+g_2(x_2,r_1,r_3),\;\;\;
\eta_{\overline{pu_3}}=(\alpha_1+\alpha_2)\overline{pu_3}+g_3(x_2,r_1,r_2),\nonumber\\[0.1em]
&\;\;\;
\eta_{\overline{u_1p}}=(\alpha_1+\alpha_2)\overline{u_1p}+h_1(x_2,r_2,r_3),\;\;\;
\eta_{\overline{u_2p}}=(\alpha_1+\alpha_2)\overline{u_2p}+h_2(r_1,r_3),\nonumber\\[0.1em]
&\;\;\;
\eta_{\overline{u_3p}}=(\alpha_1+\alpha_2)\overline{u_3p}+h_3(x_2,r_1,r_2),\nonumber\\[0.1em]
&\;\;\; \eta_{T_{11}^{{\scriptscriptstyle
(1)}}}=\alpha_2T_{11}^{{\scriptscriptstyle (1)}}-r_1\frac{\partial
q_{21}(r_2,r_3)}{\partial r_2}+q_{11}(x_2,r_2,r_3),\nonumber\\[0.1em]
&\;\;\; \eta_{T_{12}^{{\scriptscriptstyle
(1)}}}=\alpha_2T_{12}^{{\scriptscriptstyle
(1)}}-2r_{2}\frac{\partial f_{12}}{\partial
r_3}\frac{d\bar{u}_3}{dx_2}-2r_{2}\frac{\partial f_{12}}{\partial
r_3}+2f_{13}-\frac{\partial h_1}{\partial
r_2}+q_{21}(r_2,r_3)+q_{12}(x_2,r_3),\nonumber\\[0.1em]
&\;\;\; \eta_{T_{13}^{{\scriptscriptstyle
(1)}}}=\alpha_2T_{13}^{{\scriptscriptstyle (1)}}-\frac{\partial
h_1}{\partial r_3},\;\;\;\eta_{T_{21}^{{\scriptscriptstyle
(1)}}}=\alpha_2T_{21}^{{\scriptscriptstyle (1)}}
-2f_{22}\frac{d\bar{u}_1}{dx_2}-\frac{\partial h_2}{\partial
r_1}+q_{21}(r_2,r_3),\nonumber\\[0.1em]
&\;\;\; \eta_{T_{22}^{{\scriptscriptstyle
(1)}}}=\alpha_2T_{22}^{{\scriptscriptstyle
(1)}}+q_{22}(r_1,r_3),\;\;\;\eta_{T_{23}^{{\scriptscriptstyle
(1)}}}=\alpha_2T_{23}^{{\scriptscriptstyle (1)}}
-2f_{22}\frac{d\bar{u}_3}{dx_2}-2f_{22}-\frac{\partial
h_2}{\partial r_3},\nonumber\\[0.1em]
&\;\;\;\eta_{T_{31}^{{\scriptscriptstyle
(1)}}}=\alpha_2T_{31}^{{\scriptscriptstyle (1)}}
-2f_{32}\frac{d\bar{u}_1}{dx_2}-\frac{\partial h_3}{\partial
r_1}+q_{31}(x_2,r_2),\;\;\; \eta_{T_{32}^{{\scriptscriptstyle
(1)}}}=\alpha_2T_{32}^{{\scriptscriptstyle (1)}}-\frac{\partial
h_3}{\partial
r_2}+2f_{33},\nonumber\\[0.35em]
&\;\;\;
\eta_{T_{33}^{{\scriptscriptstyle (1)}}}=
\alpha_2T_{33}^{{\scriptscriptstyle (1)}}+q_{33}(x_2,r_1,r_2),\nonumber\\[0.1em]
&\;\;\; \eta_{T_{11}^{{\scriptscriptstyle
(2)}}}=\alpha_2T_{11}^{{\scriptscriptstyle
(2)}}-\big(f_{12}+f_{21}\big)\frac{d\bar{u}_1}{dx_2}
-r_2\frac{\partial f_{11}}{\partial r_3}\frac{d\bar{u}_3}{dx_2}+
r_1\frac{\partial q_{21}(r_2,r_3)}{\partial r_2}-q_{11}(x_2,r_2,r_3),\nonumber\\[0.1em]
&\;\;\; \eta_{T_{12}^{{\scriptscriptstyle
(2)}}}=\alpha_2T_{12}^{{\scriptscriptstyle
(2)}}-f_{22}\frac{d\bar{u}_1}{dx_2}+r_{2}\frac{\partial
f_{12}}{\partial r_3}\frac{d\bar{u}_3}{dx_2}+2r_{2}\frac{\partial
f_{12}}{\partial r_3}+\frac{\partial g_2}{\partial
r_1}-q_{21}(r_2,r_3)-q_{12}(x_2,r_3),\nonumber\\[0.1em]
&\;\;\; \eta_{T_{13}^{{\scriptscriptstyle
(2)}}}=\alpha_2T_{13}^{{\scriptscriptstyle
(2)}}-f_{23}\frac{d\bar{u}_1}{dx_2}-f_{12}
\frac{d\bar{u}_3}{dx_2}-2f_{12}+\frac{\partial g_3}{\partial
r_1},\nonumber\\[0.1em]
&\;\;\; \eta_{T_{21}^{{\scriptscriptstyle
(2)}}}=\alpha_2T_{21}^{{\scriptscriptstyle
(2)}}+f_{22}\frac{d\bar{u}_1}{dx_2}-r_{2}\frac{\partial
f_{21}}{\partial
r_3}\frac{d\bar{u}_3}{dx_2}+2f_{31}-\frac{\partial g_1}{\partial
x_2}+\frac{\partial g_1}{\partial r_2}-q_{21}(r_2,r_3),\nonumber\\[0.1em]
&\;\;\; \eta_{T_{22}^{{\scriptscriptstyle
(2)}}}=\alpha_2T_{22}^{{\scriptscriptstyle
(2)}}-r_{2}\frac{\partial f_{22}}{\partial
r_1}\frac{d\bar{u}_1}{dx_2}-r_{2}\frac{\partial f_{22}}{\partial
r_3}\frac{d\bar{u}_3}{dx_2}+2\big(f_{23}+f_{32}\big)-\frac{\partial
g_2}{\partial x_2}-q_{22}(r_1,r_3),\nonumber\\[0.1em]
&\;\;\; \eta_{T_{23}^{{\scriptscriptstyle
(2)}}}=\alpha_2T_{23}^{{\scriptscriptstyle
(2)}}-r_{2}\frac{\partial f_{23}}{\partial
r_1}\frac{d\bar{u}_1}{dx_2}+f_{22}\frac{d\bar{u}_3}{dx_2}+2f_{33}-\frac{\partial
g_3}{\partial x_2}+\frac{\partial g_3}{\partial
r_2},\nonumber\\[0.1em]
&\;\;\; \eta_{T_{31}^{{\scriptscriptstyle
(2)}}}=\alpha_2T_{31}^{{\scriptscriptstyle
(2)}}+f_{32}\frac{d\bar{u}_1}{dx_2}-f_{21}
\frac{d\bar{u}_3}{dx_2}-2f_{21}+\frac{\partial g_1}{\partial
r_3}-q_{31}(x_2,r_2),\nonumber\\[0.1em]
&\;\;\; \eta_{T_{32}^{{\scriptscriptstyle
(2)}}}=\alpha_2T_{32}^{{\scriptscriptstyle
(2)}}-r_{2}\frac{\partial f_{32}}{\partial
r_1}\frac{d\bar{u}_1}{dx_2}-f_{22}\frac{d\bar{u}_3}{dx_2}-2f_{22}+\frac{\partial
g_2}{\partial r_3},\nonumber\\[0.1em]
&\;\;\; \eta_{T_{33}^{{\scriptscriptstyle
(2)}}}=\alpha_2T_{33}^{{\scriptscriptstyle (2)}}-r_2\frac{\partial
f_{33}}{\partial
r_1}\frac{d\bar{u}_1}{dx_2}-\big(f_{23}+f_{32}\big)\frac{d\bar{u}_3}{dx_2}
-2\big(f_{23}+f_{32}\big)-q_{33}(x_2,r_1,r_2),
\label{160915:1053}
\end{align}
\newpage\noindent
is also compatible to the mean velocity constraint
\eqref{160915:0842}, where all $f_{ij}$, $g_i$, $h_i$, and
$q_{ij}$ are arbitrary functions only restricted by the identity
constraints [Eq.$\,$(2.6)]
\begin{equation}
\left.
\begin{aligned}
R_{ij}(\vx,\vr)=R_{ji}(\vx+\vr,-\vr),\quad
\overline{u_ip}(\vx,\vr)=\overline{pu_i}(\vx+\vr,-\vr),\\[0.0em]
T_{ij}^{{\scriptscriptstyle (1)}}(\vx,\vr)=
T_{ji}^{{\scriptscriptstyle
(2)}}(\hat{\vx},\hat{\vr})\Big|_{\hat{\vx}=\vx+\vr\,
;\,\hat{\vr}=-\vr}\, ,\hspace{1.75cm}
\end{aligned}
~~~\right\}
 \label{160915:2054}
\end{equation}
to be satisfied when generating any invariant functions for
$R_{ij}$, $\overline{u_ip}$, $\overline{pu_i}$,
$T_{ij}^{{\scriptscriptstyle (1)}}$ or
$T_{ij}^{{\scriptscriptstyle (2)}}$ from this symmetry. Note that
\eqref{160915:1053} is {\it not} the most general symmetry which
the inviscid TPC system~\eqref{160915:0858} under the mean
velocity constraint \eqref{160915:0842} can admit. It is only a
particular subgroup of a more general one not shown here. The
particular choice \eqref{160915:1053} should only give an idea as
how such a symmetry involving arbitrary functions can look like.
By specifying $\alpha_1=a_1$ and $\alpha_2=2a_1$, and by putting
all arbitrary functions in \eqref{160915:1053} to zero, this
symmetry reduces to \eqref{160915:0840}, that is,
$\mathsf{S}_1^{{\scriptscriptstyle
(1.1)}}\subset\mathsf{S}_2^{{\scriptscriptstyle (1.1)}}$ is a
subgroup of the symmetry group~\eqref{160915:1053}.

It is also important to note that although the symmetry
\eqref{160915:1053} is consistent from the outset only up to
second order in the moments,\footnote[2]{The consistency to second
order is simply due to the fact that the symmetry
\eqref{160915:1053} is only being admitted by a second order
system \eqref{160915:0858} with unclosed third-order moments
$R_{i(jk)}$ and $R_{(ik)j}$, or, respectively, expressed as
$T_{ij}^{{\scriptscriptstyle (1)}}$ and
$T_{ij}^{{\scriptscriptstyle (2)}}$ via the divergence relation
\eqref{160920:0859}. For a higher order consistency, additional
transport equations for the unclosed higher order moments need to
be considered. Hereby it does not matter whether the equations are
formulated for the $R$- or for the $T$-quantities, because if the
transformation rule for one of these quantities~is known then the
transformation rule for the other quantity can be
straightforwardly reconstructed via \eqref{160920:0859} by just
transforming along the coordinates $(\vx,\vr)$ of the underlying
symmetry.} this result can be made consistent to any order.
Because, when augmenting the inviscid TPC system
\eqref{160915:0858} by transport equations for the next
higher-order moments, one way of ensuring the stability for the
second-moment generators of \eqref{160915:1053} is simply to
enforce them as a constraint in the symmetry-finding algorithm for
the next higher order; similar as to the procedure for the mean
velocities, where \eqref{160915:0842} acted as the lower order
constraint for the symmetries of the next higher-order quantities
$R_{ij}$, $\overline{u_ip}$ and $\overline{pu_i}$ in
\eqref{160915:0858}. Due to an infinite hierarchy of equations,
this procedure is realizable at any order, since at each order
there always will be enough (unclosed) higher-order moments which
can compensate for the given constraints at lower order. And this
strategy is independent of whether one augments the inviscid TPC
system \eqref{160915:0858} by higher-order transport equations
within the two-point correlations directly, or indirectly by first
going over to the equations for the three-point correlations and
to then take the two-point limit at the end of the performed
symmetry analysis.

Although our new symmetry $\mathsf{S}_2^{{\scriptscriptstyle
(1.1)}}$ \eqref{160915:1053} is more general than the symmetry
$\mathsf{S}_1^{{\scriptscriptstyle (1.1)}}$ \eqref{160915:0840}
proposed in \cite{Oberlack06}, there is no reason to rejoice. The
problem is that we are faced with complete arbitrariness in
constructing invariant scaling laws for the second order moments
from~\eqref{160915:1053}. Due to the arbitrary and thus unknown
functions $f_{ij}$, $g_i$ and $h_i$, {\it any} arbitrary scaling
law or scaling dependency, in particular in the inhomogeneous
direction $x_2$, can be generated now by also showing full
compatibility to the velocity constraint \eqref{160915:0842}. And
in knowing that the new symmetry \eqref{160915:1053} basically
only forms a particular subgroup of a functionally much wider
group makes the situation in the problem of invariant scaling even
more worse.

As if that were not enough, the problem of arbitrariness in
invariant scaling not only extends in the direction of higher
orders, but also in the opposite direction to lower orders, namely
directly down to the initial constraints we pose. In other words,
the constraint \eqref{160915:0842} for the mean velocities itself
is not unique. Any other constraint can be posed. For example,
instead of posing linear functions for both the mean stream- and
spanwise velocity according to \eqref{160915:0842}, we will now
pose a linear function only for the streamwise velocity
$\bar{u}_1$, while for the more complex spanwise velocity
$\bar{u}_3$ we will pose a quadratic profile which, of course, as
a profile with at most three parameters, will match the DNS data
better than only a linear profile with at most two parameters.
Hence, if we pose instead of \eqref{160915:0842}, for example, the
following (closed form) constraint for the mean velocities
\begin{equation}
\bar{u}_1(x_2)=c_{11}x_2+c_{12},\quad\;\;
\bar{u}_3(x_2)=c_{31}x_2^2+c_{32}x_2+c_{33},\label{160915:1846}
\end{equation}
then the inviscid TPC system \eqref{160915:0858} being consistent
up to second order, will admit, for example, the following
symmetry group\footnote[2]{The infinitesimal generators for the
unclosed higher-order moments were not listed in
\eqref{160915:2339} since they are similar to those listed in
\eqref{160915:1053}, except for an extra quadratic term in some
components. Using the same strategy as just discussed before, also
symmetry \eqref{160915:2339} can be made consistent beyond the
second order by just keeping the result for the generators up to
second order fixed in posing them as a constraint for all higher
orders.}
\begin{align}
\mathsf{S}_3^{(1.7)}:&\;\;\; \xi_{r_1}=b_1,\;\;\;
\xi_{r_2}=0,\;\;\; \xi_{r_3}=b_3,\;\;\;\xi_{x_2}=0,\nonumber\\
&\;\;\; \eta_{R_{11}}=\beta R_{11}+\theta_{11}(x_2,r_2,r_3),\;\;\;
\eta_{R_{12}}=\beta R_{12}+\theta_{12}(x_2,r_3),\;\;\;
\eta_{R_{13}}=\beta R_{13}+\theta_{13}(x_2,r_2),\nonumber\\
&\;\;\; \eta_{R_{21}}=\beta R_{21}+\theta_{21}(r_3),\;\;\;
\eta_{R_{22}}=\beta R_{22}+\theta_{22}(r_1,r_3),\;\;\;
\eta_{R_{23}}=\beta R_{23}+\theta_{23}(r_1),\nonumber\\
&\;\;\; \eta_{R_{31}}=\beta R_{31}+\theta_{31}(x_2,r_2),\;\;\;
\eta_{R_{32}}=\beta R_{32}+\theta_{32}(x_2,r_1),\;\;\;
\eta_{R_{33}}=\beta R_{33}+\theta_{33}(x_2,r_1,r_2),\nonumber\\
&\;\;\;\eta_{\overline{pu_1}}=\beta\,\overline{pu_1}+\phi_1(x_2,r_2,r_3),\;\;\;
\eta_{\overline{pu_2}}=\beta\,\overline{pu_2}+\phi_2(x_2,r_1,r_3),\nonumber\\
&\;\;\;\eta_{\overline{pu_3}}=\beta\,\overline{pu_3}+\phi_3(x_2,r_1,r_2),
\;\;\;\eta_{\overline{u_1p}}=\beta\,\overline{u_1p}+\psi_1(x_2,r_2,r_3),\nonumber\\
&\;\;\;\eta_{\overline{u_2p}}=\beta\,\overline{u_2p}+\psi_2(r_1,r_3),
\;\;\;\eta_{\overline{u_3p}}=\beta\,\overline{u_3p}+\psi_3(x_2,r_1,r_2),
\label{160915:2339}
\end{align}
being again only a particular subgroup of a much wider and more
general group. Also here, as in symmetry
$\mathsf{S}_2^{{\scriptscriptstyle (1.1)}}$ \eqref{160915:1053},
we face again the same kind of arbitrariness in invariant scaling
for the second-order moments due to the appearance of the unknown
and thus arbitrary functions $\theta_{ij}$, $\phi_i$ and $\psi_i$,
which again are only restricted by the identity constraints
\eqref{160915:2054}. To briefly illustrate the action of these
constraints, let us consider for example the invariant function of
the diagonal component $R_{11}$ under the simplified condition
that $b_3=0$, and $\beta=0$, for which it then takes the invariant
form
\begin{equation}
R_{11}(x_2,\vr)=\theta_{11}(x_2,r_2,r_3)\cdot\frac{r_1}{b_1}+\Lambda_{11}(x_2,r_2,r_3),
\label{160915:2326}
\end{equation}
where $\Lambda_{11}$ is an arbitrary integration function. Indeed,
a quick check shows that \eqref{160915:2326} stays invariant under
the transformation
\begin{align}
\mathsf{T}_1:\;\;\; x_2^*=x_2,\;\;\; r_1^*=r_1+b_1,\;\;\;
r_2^*=r_2,\;\;\; r_3^*=r_3,\;\;\;
R_{11}^*=R_{11}+\theta_{11}(x_2,r_2,r_3),
\end{align}
induced by the symmetry generators
$\xi_{x_2}=\xi_{r_2}=\xi_{r_3}=0$, $\xi_{r_1}=b_1$ and
$\eta_{R_{11}}=\theta_{11}(x_2,r_2,r_3)$ of~\eqref{160915:2339}.
Now, the only way for \eqref{160915:2326} to satisfy the
constraint $R_{11}(x_2,\vr)=R_{11}(x_2+r_2,-\vr)$ is to restrict
the arbitrary functions $\theta_{11}$ and $\Lambda_{11}$ to the
following adapted but still invariant form\footnote[3]{The
structure of \eqref{160915:2356} is based on the fact that
$2x_2+r_2$ is an invariant under the transformation $x_2\to
x_2+r_2$ and $r_2\to -r_2$.}
\begin{equation}
R_{11}(x_2,\vr)=\widehat{\theta}_{11}(2x_2+r_2,r_2,r_3)\cdot\frac{r_1}{b_1}+
\widehat{\Lambda}_{11}(2x_2+r_2,r_2,r_3), \label{160915:2356}
\end{equation}
where $\widehat{\theta}_{11}$ and $\widehat{\Lambda}_{11}$ have to
satisfy the following conditions in their second and third
argument:
\begin{equation}
\widehat{\theta}_{11}(\,\text{\boldmath$\cdot$}\,
,-r_2,-r_3)=-\widehat{\theta}_{11}(\,\text{\boldmath$\cdot$}\,,r_2,r_3),
\quad\;\; \widehat{\Lambda}_{11}(\,\text{\boldmath$\cdot$}\,
,-r_2,-r_3)=\widehat{\Lambda}_{11}(\,\text{\boldmath$\cdot$}\,
,r_2,r_3).
\end{equation}
If we now turn into the one-point limit
($\vr\rightarrow\boldsymbol{0}$), where the two-point correlation
$R_{11}$ reduces to\linebreak the diagonal Reynolds stress
component $\tau_{11}$ (up to an error $\mathcal{O}(Re^{-1/2})$;
see first footnote on p.$\,$\pageref{footnote-p2})\linebreak and
where we assume that the free functions $\widehat{\theta}_{11}$
and $ \widehat{\Lambda}_{11}$ behave smoothly in this limit, then
we will obtain, according to \eqref{160915:2356}, the following
fully arbitrary result
\begin{equation}
\lim_{\vr\to\boldsymbol{0}}R_{11}(x_2,\vr)=\widehat{\Lambda}_{11}(2x_2)=:\tau_{11}(x_2),
\label{160920:2145}
\end{equation}
which gives no information at all as how the Reynolds stress
tensor $\tau_{11}$ should scale if we assume a linear and a
quadratic mean velocity profile for $\bar{u}_1$ and $\bar{u}_3$,
respectively, according to the given constraint
\eqref{160915:1846}. Any function $\widehat{\Lambda}_{11}$ within
any region of $x_2$ can thus be chosen such that the numerical or
experimental results are matched satisfactorily.

Even if we relax the condition of zero to a non-zero scaling
$\beta\neq 0$, the problem of arbitrariness remains. Instead of
\eqref{160915:2326}, the invariant function now takes the form
\begin{equation}
R_{11}(x_2,\vr)=-\frac{1}{\beta}\cdot\theta_{11}(x_2,r_2,r_3)+e^{\beta
r_1/b_1}\cdot\Gamma_{11}(x_2,r_2,r_3), \label{160920:1948}
\end{equation}
which indeed stays invariant under the 1-parametric $(\epsilon)$
group
 transformation
\begin{align}
\!\!\mathsf{T}_2:&\;\: x_2^*=x_2,\;\: r_1^*=r_1+b_1\epsilon,\;\:
r_2^*=r_2,\;\: r_3^*=r_3,\;\:
R_{11}^*=e^{\beta\epsilon}R_{11}+\frac{e^{\beta\epsilon}-1}{\beta}\cdot\theta_{11}(x_2,r_2,r_3),
\label{160920:2130}
\end{align}
induced by the generators $\xi_{x_2}=\xi_{r_2}=\xi_{r_3}=0$,
$\xi_{r_1}=b_1$ and $\eta_{R_{11}}=\beta
R_{11}+\theta_{11}(x_2,r_2,r_3)$ of symmetry \eqref{160915:2339}.
Again, for \eqref{160920:1948} to satisfy the constraint
$R_{11}(x_2,\vr)=R_{11}(x_2+r_2,-\vr)$, the arbitrary integration
function $\Gamma_{11}$ has to be turned down to zero, while
$\theta_{11}$ needs to be restricted~to$\,$\footnote[2]{The
restriction from \eqref{160920:1948} to \eqref{160920:2112} did
not change the considered invariance property, that is, function
\eqref{160920:2112} is still invariant under the considered
transformation \eqref{160920:2130}.}
\begin{equation}
R_{11}(x_2,\vr)=-\frac{1}{\beta}\cdot\widehat{\theta}_{11}(2x_2+r_2,r_2,r_3),
\label{160920:2112}
\end{equation}
where $\widehat{\theta}_{11}$ has to be a symmetric function now
in its second and third argument:
\begin{equation}
\widehat{\theta}_{11}(\,\text{\boldmath$\cdot$}\,
,-r_2,-r_3)=\widehat{\theta}_{11}(\,\text{\boldmath$\cdot$}\,,r_2,r_3).
\end{equation}
Hence, as before in \eqref{160920:2145}, the one-point limit of
\eqref{160920:2112} leads again to a fully arbitrary result for
the invariant Reynolds stress
\begin{equation}
\lim_{\vr\to\boldsymbol{0}}R_{11}(x_2,\vr)=
-\frac{1}{\beta}\cdot\widehat{\theta}_{11}(2x_2)=:\tau_{11}(x_2).
\label{160920:2148}
\end{equation}

Coming back to the initial ansatz \eqref{160915:1846}, it is clear
that for the mean streamwise velocity field $\bar{u}_1$ we could
have also chosen a different functional dependency than the linear
one proposed in \cite{Oberlack06}. Obviously, scaling the
corresponding DNS results [Fig.$\,$3, p.$\,$393] by eye it is
reasonable to assume a linear scaling law for $\bar{u}_1$ in the
range $x_2\sim 0.2$-$0.6$ for $Ro>2.5$. But a systematic group
analysis does not uniquely predict this behavior. It can also be a
weak (non-linear) power law, which, for example, could serve as an
alternative constraint for $\bar{u}_1$ in \eqref{160915:1846}.
Analytically with group theory alone, it is not possible to tell
how the mean velocity profile $\bar{u}_1$ really scales. In
particular, it is not clear at all how the scaling behavior will
change with ever increasing rotation rates. Maybe the ``linear
scaling" weakens and gets less obvious~by~eye.\linebreak

\vspace{0.0em}
\begin{center}
\textbf{--- Conclusion ---}
\end{center}
As a result of this section we can conclude that a Lie-group
symmetry analysis on the unclosed TPC equation \eqref{160915:0858}
cannot analytically predict its scaling behavior {\it a priori}.
For that, modelling procedures and exogenous information from
numerical simulations or physical experiments are needed to get
further insights. Ultimately this just reflects the closure
problem of turbulence, which, as we have clearly demonstrated in
this section, cannot be solved or bypassed by the method of
Lie-groups alone, as misleadingly and continually claimed by
Oberlack~{\it et~al.} also again in their latest contribution
\cite{Oberlack16C}.

Neither by augmenting the unclosed TPC equation
\eqref{160915:0858} with transport equations for its higher-order
moments, nor by including the three-point correlation equations,
the problem of arbitrary scaling cannot be circumvented. Every
systematic Lie-group symmetry analysis will always lead to a
sufficient number of free functions such that for every turbulent
flow quantity any kind of invariant scaling law can be generated,
which, within a trial and error procedure, can be always chosen
such as to fit any numerical or experimental data
adequately.\pagebreak[4]

\newgeometry{left=2.5cm,right=2.5cm,top=2.5cm,bottom=2.0cm,headsep=1em}

Also when considering the strategy as initially proposed in
\cite{Oberlack10} and then as later applied in
\cite{Oberlack13.1}, \cite{Oberlack14}, \cite{Oberlack14.1} and
\cite{Oberlack15Rev}, namely to formally consider the infinite
hierarchy of multi-point equations, the problem of arbitrariness
in invariant scaling remains, independent of whether the pure
fluctuating $R$- or the instantaneous $H$-approach
in~\cite{Oberlack10}\linebreak is used.\footnote[2]{To note is
that in particular the later studies \cite{Oberlack10},
\cite{Oberlack13.1}, \cite{Oberlack14.1} and \cite{Oberlack14}
also suffer from the additional problem that new unphysical
symmetries are generated, which in turn violate the classical
principle of cause and effect. For more details, we refer to our
other comments and reviews,
\cite{Frewer14.1,Frewer15.1,Frewer16.1,Frewer16.2,Frewer16,Frewer16.3},
and to our reactions in
\cite{Frewer.X1,Frewer.X2,Frewer.X3,Frewer.X4}.} The key problem
here is that the infinite multi-point system is a forward
recursive hierarchy \citep{Frewer15.0,Frewer15.0x}, in which the
(unclosed) higher-order $n$-point correlations can only be
obtained by the next higher $(n+1)$-point correlation equation,
but which by itself is again unclosed, rendering thus the infinite
hierarchy to an unclosed system admitting arbitrary symmetries,
because as with each higher order new arbitrary functions in the
symmetry finding process will appear. Particularly in the
instantaneous $H$-approach the closure problem can be experienced
directly when performing a symmetry analysis on the infinite
hierarchy of multi-point equations. Because, since the hierarchy
in this representation is linear, it naturally admits the symmetry
of linear superposition, thus giving raise to a symmetry which is
unclosed \emph{per se}: Any solution solving the underlying
(unclosed) system of multi-point equations up to a certain order
can be added or superposed to an already given symmetry to obtain
a new symmetry. Incrementally one can therefore now improve the
symmetry for any turbulent flow quantity such as to obtain an
invariant scaling law that will adequately fit the data. Hence,
the systematic Lie-group symmetry approach degenerates down to a
non-predictive incremental trial-and-error method. For more
details on this issue, see e.g. the instructive example
in~\cite{Frewer16.3}.\linebreak \indent $\,$ Hence, in this sense
Lie-group theory offers no answer, nor does it give any
prediction\linebreak {\it a priori} in how turbulence should
scale. As already said, this failure simply reflects the classical
closure problem of turbulence, which, also with the powerful and
appealing Lie-group symmetry method, cannot be solved or bypassed
analytically. However, using this method to nevertheless
systematically generate such invariant scaling laws would be the
same as guessing it, and if one knows what to expect {\it a
posteriori} then, of course, one can manually arrange everything
backwards and pretend that theory is predicting these results. But
such an approach has nothing to do with science
\citep{Frewer14.2}.

\vspace{0em}
\section{List of inconsistencies and incorrect statements in
Oberlack \emph{et al.}~(2006)}

This section will reveal all inconsistent and incorrect
information that can be found in Sec.$\,$2.3 [pp.$\,$388-392] of
\cite{Oberlack06}. They will be listed and discussed in the order
as they appear in the text.

$\!$\emph{\textbf{(1):}} It is claimed that ``for physical reasons
the translation invariance of $\tilde{r}_i$ is not
meaningful",\linebreak with the argument that ``since
$\tilde{R}_{ij}$ reaches its finite maximum at $|\tilde{\vr}|=0$
and tends to zero for $|\tilde{\vr}|\to\infty$, a shift in the
correlation space cannot be a new solution" [p.$\,$390]. If one
would strictly follow this argument, then consequently it also has
to apply to the translation invariance of $\tilde{x}_2$. Because,
since $\tilde{R}_{ij}$ has one of its minima always at
$\tilde{x}_2=\pm 1$ due to the no-slip condition at the walls, a
shift in wall-normal direction thus cannot be a new solution, too.
Hence, according to \cite{Oberlack06} also the translation
invariance of $\tilde{x}_2$ should not be physically meaningful,
and thus in the same way as the translation parameters $a_2$-$a_4$
for the correlations lengths $\tilde{r}_i$ were put to zero, so
has the translation parameter $a_5$ in the wall-normal
$\tilde{x}_2$-direction be put to zero, if one would strictly
follow the reasoning in \cite{Oberlack06}. However, when putting
$a_5=0$ has a significant negative effect on the results obtained
for the similarity variables $\eta_i$~[Eq.$\,$(2.23a)]. Because,
in or near the channel center plane $\tilde{x}_2=0$ we would have
infinitely large values for all three independent similarity
variables $\eta_i$, which again would yield the unphysical result
of zero correlations in that region, since by definition
$\tilde{R}_{ij}$ tends to zero for $|\veta|\to\infty$.\newpage

\newgeometry{left=2.5cm,right=2.5cm,top=2.5cm,bottom=2.0cm,headsep=1em}

\begin{figure}[t]
\centering
\begin{minipage}[c]{.48\linewidth}
\FigureXYLabel{\includegraphics[width=.91\textwidth]{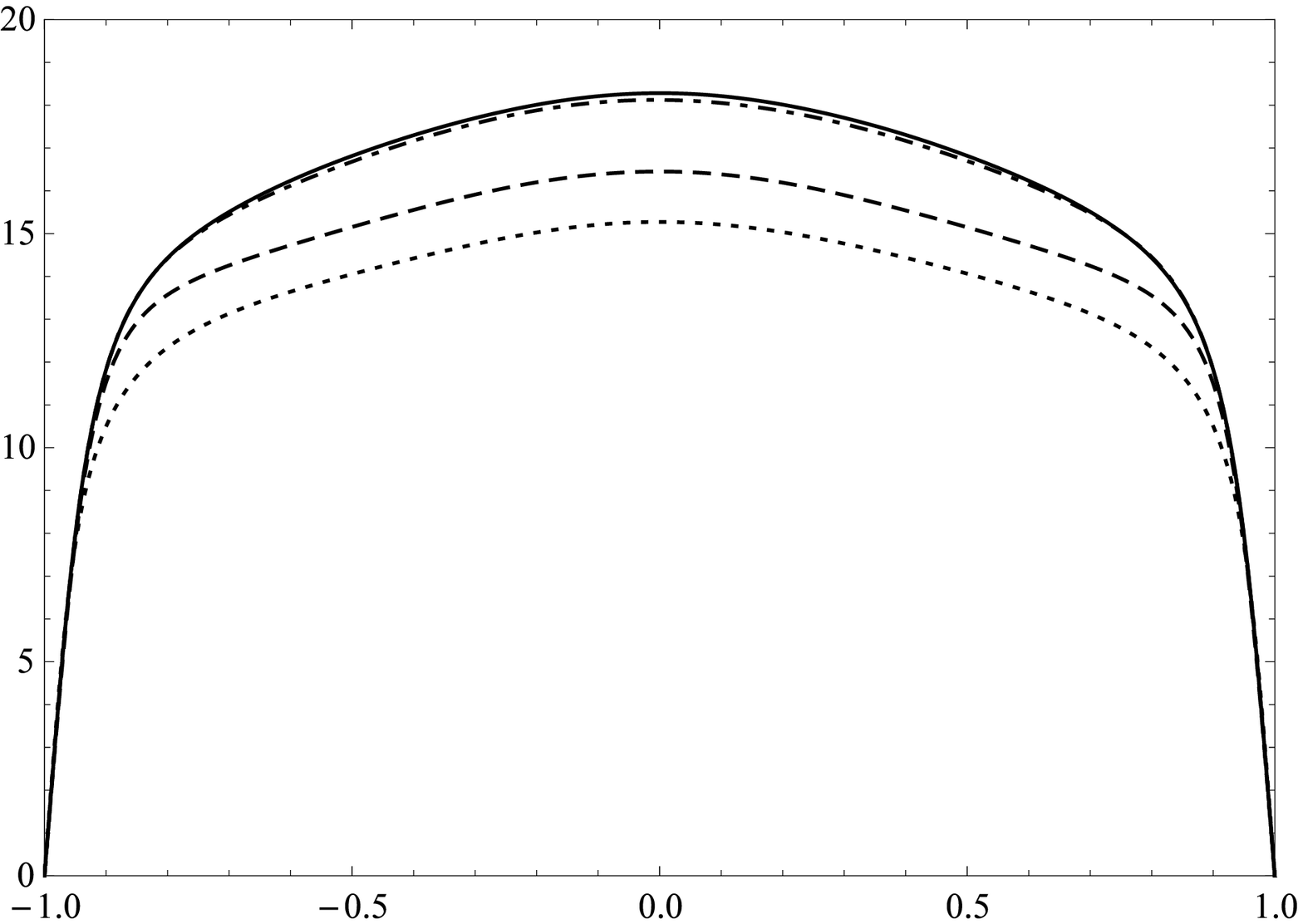}}
{${\scriptstyle\hspace{0.5cm} x_2}$}{-1mm}{\begin{rotate}{0}
$\hspace{-0.2cm}{\scriptstyle \bar{u}_1^+}$\end{rotate}}{1.5mm}
\end{minipage}
\hfill
\begin{minipage}[c]{.48\linewidth}
\FigureXYLabel{\includegraphics[width=.91\textwidth]{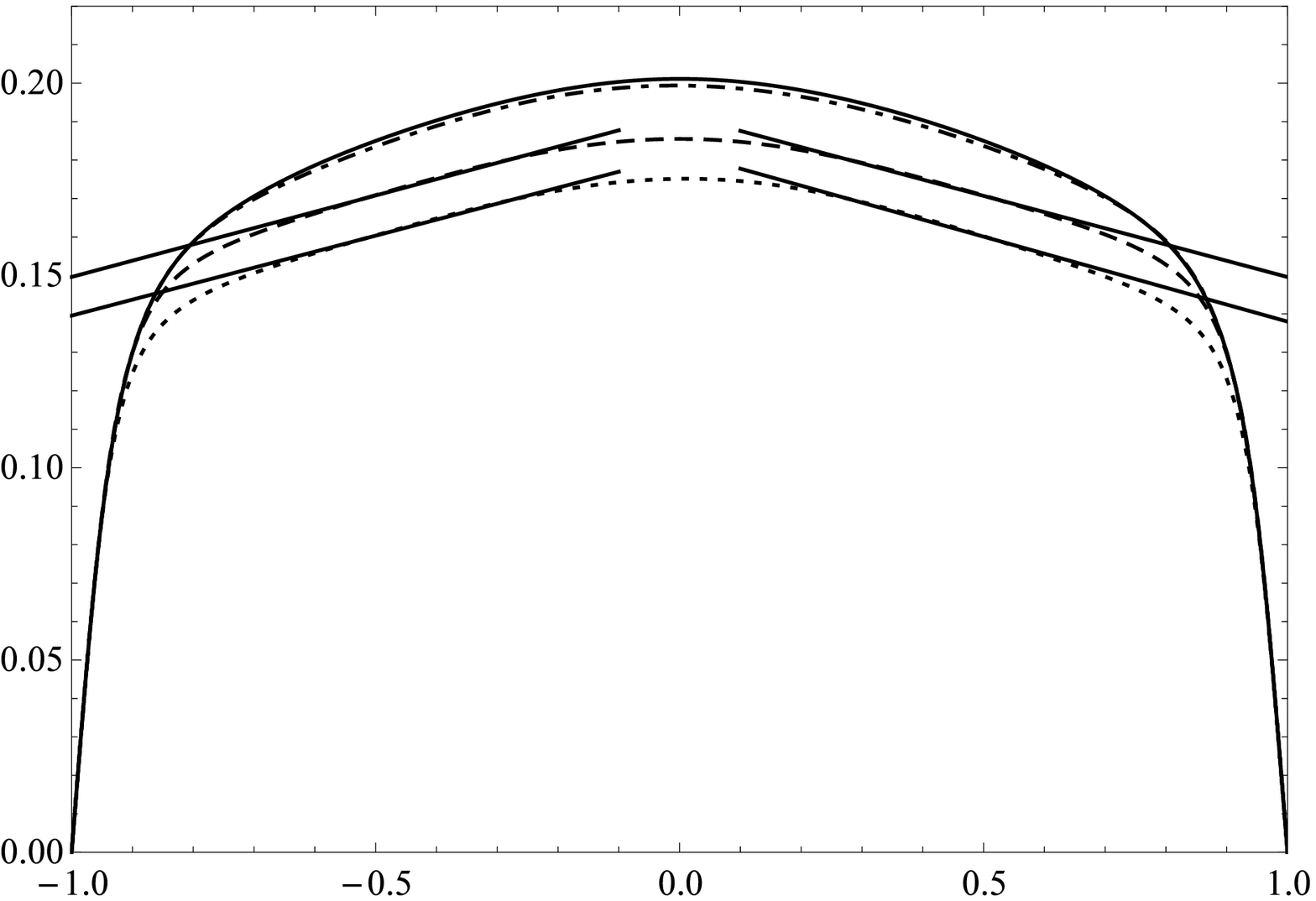}}
{${\scriptstyle\hspace{0.65cm} x_2}$}{-1mm}{\begin{rotate}{0}
$\hspace{-0.2cm}{\scriptstyle \bar{u}_1}$\end{rotate}}{1.5mm}
\end{minipage}
\caption{DNS result for the mean streamwise velocity field at
$Re_\tau=180$ for different rotation numbers $Ro=0; 2.5; 6.5$ and
$10$ (from top to bottom). The numerical simulation was repeated
with the same code \citep{Lundbladh92} for the same parameters,
domain sizes and resolution as chosen in~\cite{Oberlack06}. The
left plot exactly corresponds to Fig.$\,$3 [p.$\,$393] in
\cite{Oberlack06} and displays the velocity profile normalized on
$u_\tau$, which for all of the considered rotation rates takes the
same unchanged value $u_\tau\sim 0.011$ at $Re_\tau=180$. The
right plot displays the dimensionalized or non-normalized velocity
profile including the best fit of the invariant linear scaling law
$\bar{u}_1=m\cdot x_2+b$ according to \eqref{160922:0904}. Already
with a normal sense of proportion one can see that the proposed
linear scaling cannot meet the conditions as claimed in
\cite{Oberlack06} for the range $|x_2|=0.2$-$0.6$: The slope $m$
does not proportionally scale with the rotation rate
$Ro\sim\Omega$, but rather stays invariant at a value $|m|=0.042$,
while the constant~$b$ contrarily exhibits a strong dependency on
$\Omega$: Increasing the rotation rate from $Ro=6.5$ to $10$, the
constant already decreases by more than $5\%$, from $b=0.192$ down
to $0.181$.}
\label{fig1}
\end{figure}

$\!$\emph{\textbf{(2):}} Since Eqs.$\,$(2.21a)-(2.21b) result from
Eq.$\,$(2.14) which by itself does not show any explicit
dependence on the rotation parameter $\Omega$, it is correct that
all group and integration constants appearing in
Eqs.$\,$(2.21a)-(2.21b) do not depend on $\Omega$. The dependence
on the rotation rate only enters when transforming this result
back to its original variables according to
Eqs.$\,$(2.13a)-(2.13c). Along with the claim that ``the function
$\gamma$ behaves as $\gamma\sim 1/\Omega$" [p.$\,$390], this
transformation then yields the invariant result
[Eqs.$\,$(2.22a)-(2.22b)]
\begin{equation}
\bar{u}_1\sim C_1\Omega\, x_2+
\underbrace{C_1a_5/a_1-c_1/a_1\vphantom{g_{g_{g_g}}}}_{=:\,
B_1},\quad\; \bar{u}_3\sim C_3\Omega\, x_2+
\underbrace{C_3a_5/a_1-c_3/a_1\vphantom{g_{g_{g_g}}}}_{=:\ B_3},
\label{160922:0904}
\end{equation}
with the conclusion that ``only the slope of the linear scaling
laws depends on the rotation rate" [p.$\,$390]. But such a
parametric scaling is inconsistent to the DNS results shown in
Fig.$\,$3 in \cite{Oberlack06}. Because, when fitting a linear law
$\bar{u}_1=m\cdot x_2+b$ for the streamwise velocity field to the
DNS data, we obtain, as can be seen from Figure \ref{fig1}, a
contrary result to~\eqref{160922:0904}: (i) For the lowest chosen
rotation rate $Ro=2.5$ no convincing linear scaling over a longer
range can be detected. (ii) When increasing the rotation rate from
$Ro=6.5$ to $10$, the slope $m$ of the linear law does not
proportionally increase along as prescribed by
\eqref{160922:0904}; instead it rather remains constant. (iii) The
claim that ``the two additive constants appearing in the scaling
laws (2.22a) and (2.22b) do not depend on $\Omega$" [p.$\,$390]
cannot be confirmed; the DNS results in
Figure~\ref{fig1}\linebreak clearly show the opposite, namely a
fairly strong dependence of $b=B_1$ on the rotation rate $\Omega$.

Important to note here is that for Figure~\ref{fig1} we have rerun
the DNS with the same code \citep{Lundbladh92} for the same
parameters, domain sizes and resolution as chosen in
\cite{Oberlack06}, with the only aim to check the internal
consistency of its part on group-analysis. We did not check the
consistency of the DNS results themselves as presented in
\cite{Oberlack06}, which would be a study in itself, in particular
as these results were criticized by \cite{Recktenwald09} as not
being reliable due to the prescription of periodic boundary
conditions on a too small computational domain in the spanwise
direction.\newpage

\indent $\!$\emph{\textbf{(3):}} The result Eq.$\,$(2.23b) for the
two-point velocity correlation $\tilde{R}_{ij}$ is {\it not}
consistent with the underlying one-point momentum equation
equation Eq.$\,$(2.1a) in the limit of large Reynolds numbers.
Because, when taking the one-point limit of $\tilde{R}_{ij}$
[Eq.$\,$(2.23b)]
\begin{align}
\lim_{\vr\to\boldsymbol{0}}\tilde{R}_{ij}(\tilde{x}_2,\tilde{\vr})
\quad &=\quad\lim_{\vr\to\boldsymbol{0}}
F_{ij}(\veta)\cdot(\tilde{x}_2+a_5/a_1)^2\nonumber\\
=\tilde{\tau}_{ij}(\tilde{x}_2)+\mathcal{O}(Re^{-1/2})\quad
&\hspace{0.25cm}\bigg|\hspace{0.45cm} = F_{ij}(0)\cdot
(\tilde{x}_2+a_5/a_1)^2,
\end{align}
we obtain a quadratic law for all Reynolds stresses (up to an
error $\mathcal{O}(Re^{-1/2})$, however which becomes negligibly
small in the limit of large Reynolds numbers; see first footnote
on p.$\,$\pageref{footnote-p2}):
\begin{equation}
\tau_{ij}(x_2)=\Omega^2\cdot C_{ij}\cdot
\Big(x_2+\gamma(\Omega)\cdot a_5/a_1\Big)^2, \label{160923:1012}
\end{equation}
where $C_{ij}=F_{ij}(0)$ is the arbitrary integration constant.
However, such a law does not solve the inviscid ($\nu=0$) momentum
equation Eq.$\,$(2.1a) \vspace{-0.5em}
\begin{equation}
0=K-\frac{d \tau_{12}}{dx_2},\label{160923:1029}
\end{equation}
when taking the key assumption of \cite{Oberlack06} that
\eqref{160923:1012} is valid ``in regions sufficiently far from
solid walls, [where] the viscous terms may be neglected to leading
order" [p.$\,$385]. The quantity $K$ denotes the {\it constant}
mean streamwise pressure gradient $K\sim -\partial
\bar{p}/\partial x_1$ that drives the flow. As also can be clearly
seen from the DNS result in Fig.$\,$6 [p.$\,$395], the shear
stress $\tau_{12}=\overline{u_1u_2}$ follows a linear law away
from the walls consistent with equation \eqref{160923:1029}, and
not according to the quadratic law \eqref{160923:1012} as
incorrectly proposed in \cite{Oberlack06}.

$\!$\emph{\textbf{(4):}} The result Eq.$\,$(2.26) [p.$\,$391] is
incorrect. A substantial factor is missing, which, when included,
invalidates the conclusion in \cite{Oberlack06} that ``relation
(2.26) gives raise to a new symmetry transformation" [p.$\,$391].
Also the claim that the validity of this symmetry transformation
[Eq.$\,$(2.27)] ``can be verified by substituting (2.27) into
(2.14) after the similarity coordinate (2.25) and the linear
profiles (2.21a) and (2.21b) have been employed" [p.$\,$391],
cannot be confirmed, neither with the incorrect relation as given
by Eq.$\,$(2.26), nor with the correct relation
\eqref{160923:1245}, which will be derived now: Using the
invariant result of $\tilde{R}_{ij}$ [Eq.$\,$(2.23b)], the
identity
$\tilde{R}_{ij}(\tilde{x}_2,\tilde{\vr})=\tilde{R}_{ji}(\tilde{x}_2+\tilde{r}_2,-\tilde{\vr})$
[Eq.$\,$(2.6)] and the abbreviation $\tilde{x}_2^\prime
=\tilde{x}_2+a_5/a_1$ [Eq.$\,$(2.24)], then the correct derivation
of relation Eq.$\,$(2.26) reads
\begin{align}
\tilde{R}_{ij}(\tilde{x}_2,\tilde{\vr})&=F_{ij}(\veta)\cdot
\left(\tilde{x}_2+\frac{a_5}{a_1}\right)^2\nonumber\\
&= F_{ij}\left(\frac{\tilde{\vr}}{\tilde{x}_2^\prime}\right)\cdot
\tilde{x}_2^{\prime\, 2}\,\equiv\,
F_{ji}\left(\frac{-\tilde{\vr}}{\tilde{x}_2^\prime+\tilde{r}_2}\right)\cdot
\Big(\tilde{x}_2^\prime+\tilde{r}_2\Big)^2\nonumber\\
&\hspace{2.91cm}=F_{ji}\left(\frac{-\tilde{\vr}}{\tilde{x}_2^\prime\cdot
\big(1+\tilde{r}_2/\tilde{x}_2^\prime\big)}\right)\cdot
\tilde{x}_2^{\prime\, 2}\cdot
\Big(1+\tilde{r}_2/\tilde{x}_2^\prime\Big)^2\nonumber\\
&= F_{ji}\left(-\frac{\veta}{1+\eta_2}\right)\cdot
\tilde{x}_2^{\prime\, 2}\cdot \big(1+\eta_2\big)^2,
\end{align}
which then, since
$\tilde{R}_{ij}(\tilde{x}_2,\tilde{\vr})=F_{ij}(\veta)\cdot
\tilde{x}_2^{\prime\, 2}$ [Eq.$\,$(2.23b)], finally leads to the
different result
\begin{equation}
F_{ij}(\veta)=F_{ji}\left(-\frac{\veta}{1+\eta_2}\right)\cdot
\big(1+\eta_2\big)^2.\label{160923:1245}
\end{equation}

\vspace{-1.6em}
\bibliographystyle{jfm}
\bibliography{BibDaten}

\begin{thebibliography}{26}
\expandafter\ifx\csname natexlab\endcsname\relax\def\natexlab#1{#1}\fi

\bibitem[Avsarkisov {\em et~al.\/}(2014)Avsarkisov, Oberlack \&
  Hoyas]{Oberlack14}
{\sc Avsarkisov, V., Oberlack, M. \& Hoyas, S.} 2014 New scaling laws for
  turbulent {P}oiseuille flow with wall transpiration. {\em J.~Fluid~Mech.\/}
  {\bf 746}, 99--122.

\bibitem[Frewer(2015{\natexlab{{\em a\/}}})]{Frewer15.0}
{\sc Frewer, M.} 2015{\natexlab{{\em a\/}}} {A}n example elucidating the
  mathematical situation in the statistical non-uniqueness problem of
  turbulence. {\em \href{http://arxiv.org/abs/1508.06962}{arXiv:1508.06962}\/}
  {\bf $\!\!$}.

\bibitem[Frewer(2015{\natexlab{{\em b\/}}})]{Frewer15.0x}
{\sc Frewer, M.} 2015{\natexlab{{\em b\/}}} {A}pplication of {L}ie-group
  symmetry analysis to an infinite hierarchy of differential equations at the
  example of first order {O}{D}{E}s. {\em
  \href{http://arxiv.org/abs/1511.00002}{arXiv:1511.00002}\/} {\bf $\!\!$}.

\bibitem[Frewer(2015{\natexlab{{\em c\/}}})]{Frewer.X1}
{\sc Frewer, M.} 2015{\natexlab{{\em c\/}}} {O}n a remark from {J}ohn von
  {N}eumann applicable to the symmetry induced turbulent scaling laws generated
  by the new theory of {O}berlack et al. {\em
  \href{https://www.researchgate.net/publication/286732368}{ResearchGate}\/}
  {\bf \!\!\!\phantom{,}}, 1--3.

\bibitem[Frewer \& Khujadze(2016{\natexlab{{\em a\/}}})]{Frewer16}
{\sc Frewer, M. \& Khujadze, G.} 2016{\natexlab{{\em a\/}}} {C}omments on
  {J}anocha {\it et al.} {L}ie symmetry analysis of the {H}opf
  functional-differential equation". {\em
  \href{http://www.mdpi.com/2073-8994/8/4/23}{Symmetry}\/} {\bf 8}~(4), 23.

\bibitem[Frewer \& Khujadze(2016{\natexlab{{\em b\/}}})]{Frewer.X4}
{\sc Frewer, M. \& Khujadze, G.} 2016{\natexlab{{\em b\/}}} An example of how a
  methodological mistake aggravates erroneous results when only correcting the
  results and not the method itself. {\em
  \href{https://www.researchgate.net/publication/301553065}{ResearchGate}\/}
  {\bf \!\!\!\phantom{,}}, 1--10.

\bibitem[Frewer {\em et~al.\/}(2014{\natexlab{{\em a\/}}})Frewer, Khujadze \&
  Foysi]{Frewer14.1}
{\sc Frewer, M., Khujadze, G. \& Foysi, H.} 2014{\natexlab{{\em a\/}}} On the
  physical inconsistency of a new statistical scaling symmetry in
  incompressible {N}avier-{S}tokes turbulence. {\em
  \href{http://arxiv.org/abs/1412.3061}{arXiv:1412.3061}\/} {\bf $\!\!$}.

\bibitem[Frewer {\em et~al.\/}(2014{\natexlab{{\em b\/}}})Frewer, Khujadze \&
  Foysi]{Frewer14.2}
{\sc Frewer, M., Khujadze, G. \& Foysi, H.} 2014{\natexlab{{\em b\/}}}
  \phantom{x}$\!\!${I}s the log-law a first principle result from
  {L}ie-\linebreak group invariance analysis? {A} comment on the {A}rticle by
  {O}berlack (2001). {\em
  \href{http://arxiv.org/abs/1412.3069}{$\!$arXiv:1412.3069}\/} {\bf $\!\!$}.

\bibitem[Frewer {\em et~al.\/}(2015{\natexlab{{\em a\/}}})Frewer, Khujadze \&
  Foysi]{Frewer15.1}
{\sc Frewer, M., Khujadze, G. \& Foysi, H.} 2015{\natexlab{{\em a\/}}} Comment
  on ``{S}tatistical symmetries of the {L}undgren-{M}onin-{N}ovikov hierarchy".
  {\em
  \href{http://journals.aps.org/pre/abstract/10.1103/PhysRevE.92.067001}{Phys.~Rev.~E}\/}
  {\bf 92}, 067001.

\bibitem[Frewer {\em et~al.\/}(2015{\natexlab{{\em b\/}}})Frewer, Khujadze \&
  Foysi]{Frewer.X2}
{\sc Frewer, M., Khujadze, G. \& Foysi, H.} 2015{\natexlab{{\em b\/}}}
  Objections to a {R}eply of {O}berlack et al. {\em
  \href{https://www.researchgate.net/publication/287996194}{ResearchGate}\/}
  {\bf \!\!\!\phantom{,}}, 1--9.

\bibitem[Frewer {\em et~al.\/}(2016{\natexlab{{\em a\/}}})Frewer, Khujadze \&
  Foysi]{Frewer16.1}
{\sc Frewer, M., Khujadze, G.$\!$ \& Foysi, H.$\!$} 2016{\natexlab{{\em a\/}}}
  A note on the notion ``statistical symmetry". {\em
  \href{http://arxiv.org/abs/1602.08039}{arXiv:1602.08039}\/} {\bf $\!\!$}.

\bibitem[Frewer {\em et~al.\/}(2016{\natexlab{{\em b\/}}})Frewer, Khujadze \&
  Foysi]{Frewer16.2}
{\sc Frewer, M., Khujadze, G. \& Foysi, H.} 2016{\natexlab{{\em b\/}}}
  \phantom{2}$\!\!\!${C}omment on ``{A}pplication of the extended {L}ie group
  analysis to the {H}opf functional formulation of the {B}urgers equation".
  {\em
  \href{http://scitation.aip.org/content/aip/journal/jmp/57/3/10.1063/1.4940357}{J.~Math.~Phys.}\/}
  {\bf 57}, 034102.

\bibitem[Frewer {\em et~al.\/}(2016{\natexlab{{\em c\/}}})Frewer, Khujadze \&
  Foysi]{Frewer.X3}
{\sc Frewer, M., Khujadze, G. \& Foysi, H.} 2016{\natexlab{{\em c\/}}}
  \phantom{3}$\!\!\!${O}n a new technical error in a further {R}eply by
  {O}berlack et al. and its far-reaching effect on their original study. {\em
  \href{https://www.researchgate.net/publication/299368809}{ResearchGate}\/}
  {\bf \!\!\!\phantom{,}}, 1--6.

\bibitem[Khujadze \& Frewer(2016)]{Frewer16.3}
{\sc Khujadze, G. \& Frewer, M.} 2016 Revisiting the {L}ie-group symmetry
  method for turbulent channel flow with wall transpiration. {\em
  \href{https://arxiv.org/abs/1606.08396}{arXiv:1606.08396}\/} {\bf $\!\!$}.

\bibitem[Khujadze \& Oberlack(2004)]{Khujadze04}
{\sc Khujadze, G. \& Oberlack, M.} 2004 D{N}{S} and scaling laws from new
  symmetries of {Z}{P}{G} turbulent boundary layer flow. {\em
  Theor.~Comp.~Fluid~Dyn.\/} {\bf 18}, 391--411.

\bibitem[Lundbladh {\em et~al.\/}(1992)Lundbladh, Henningson \&
  Johansson]{Lundbladh92}
{\sc Lundbladh, A., Henningson, D.~S. \& Johansson, A.~V.} 1992 An efficient
  spectral integration method for the solution of the {N}avier-{S}tokes
  equations. {\em Tech.~Rep. FFA-TN 1992-28,\/} Aeronautical Research Institute
  of Sweden, Bromma.

\bibitem[Oberlack(2002)]{Oberlack02B}
{\sc Oberlack, M.} 2002 Symmetries and invariant solutions of turbulent flows
  and their implications for turbulence modelling. In {\em Theories of
  Turbulence\/} (ed. M.~Oberlack \& F.~H. Busse), pp. 301--366. Springer.

\bibitem[Oberlack {\em et~al.\/}(2006)Oberlack, Cabot, Reif \&
  Weller]{Oberlack06}
{\sc Oberlack, M., Cabot, W., Reif, B. A.~Pettersson \& Weller, T.} 2006 Group
  analysis, direct numerical simulation and modelling of a turbulent channel
  flow with streamwise rotation. {\em J.~Fluid~Mech.\/} {\bf 562}, 383--403.

\bibitem[Oberlack \& Guenther(2003)]{Oberlack03}
{\sc Oberlack, M. \& Guenther, S.} 2003 Shear-free turbulent diffusion -
  classical and new scaling laws. {\em Fluid~Dyn.~Res.\/} {\bf 33}, 453--476.

\bibitem[Oberlack \& Rosteck(2010)]{Oberlack10}
{\sc Oberlack, M. \& Rosteck, A.} 2010 New statistical symmetries of the
  multi-point equations and its importance for turbulent scaling laws. {\em
  Discrete~Continuous~Dyn.~Syst.\/} {\bf Ser. S 3}, 451--471.

\bibitem[Oberlack \& Rosteck(2016)]{Oberlack16C}
{\sc Oberlack, M. \& Rosteck, A.} 2016 Circumnavigating the closure problem of
  turbulence.\linebreak {A} {L}ie symmetry approach. {\it $24^\text{th}$
  International Congress of Theoretical and Applied Mechanics}, 21-26 August
  2016, Montreal, Canada.
  \href{https://myictam2016.zerista.com/event/member/278351}{TS.FM14-1.01}.

\bibitem[Oberlack {\em et~al.\/}(2015)Oberlack, Wac{\l}awczyk, Rosteck \&
  Avsarkisov]{Oberlack15Rev}
{\sc Oberlack, M., Wac{\l}awczyk, M., Rosteck, A. \& Avsarkisov, V.} 2015
  Symmetries and their importance for statistical turbulence theory. {\em
  Mech.~Eng.~Rev.\/} {\bf 2}~(2), 15--00157.

\bibitem[Oberlack \& Zieleniewicz(2013)]{Oberlack13.1}
{\sc Oberlack, M. \& Zieleniewicz, A.} 2013 Statistical symmetries and its
  impact on new decay modes and integral invariants of decaying turbulence.
  {\em Journal~of~Turbulence\/} {\bf 14}~(2), 4--22.

\bibitem[Recktenwald {\em et~al.\/}(2009)Recktenwald, Alkishriwi \&
  Schr{\"o}der]{Recktenwald09}
{\sc Recktenwald, I., Alkishriwi, N. \& Schr{\"o}der, W.} 2009
  P{I}{V}-{L}{E}{S} analysis of channel flow rotating about the streamwise
  axis. {\em Eur.~J.~Mech.~B/Fluids\/} {\bf 28}~(5), 677--688.

\bibitem[Vu {\em et~al.\/}(2012)Vu, Jefferson \& Carminati]{Vu12}
{\sc Vu, K.~T., Jefferson, G.~F. \& Carminati, J.} 2012 Finding higher
  symmetries of differential equations using the {MAPLE} package {DESOLVII}.
  {\em Comp.~Phys.~Comm.\/} {\bf 183}~(4), 1044--1054.

\bibitem[Wac{\l}awczyk {\em et~al.\/}(2014)Wac{\l}awczyk, Staffolani, Oberlack,
  Rosteck, Wilczek \& Friedrich]{Oberlack14.1}
{\sc Wac{\l}awczyk, M., Staffolani, N., Oberlack, M., Rosteck, A., Wilczek, M.
  \& Friedrich, R.} 2014 Statistical symmetries of the
  {L}undgren-{M}onin-{N}ovikov hierarchy. {\em Phys.~Rev.~E\/} {\bf 90}~(1),
  013022.

\end{thebibliography}

\end{document}